\documentclass[runningheads]{llncs}
\usepackage{graphicx,color,booktabs,multirow}
\usepackage{amssymb}
\usepackage{amsmath,subfigure}

\begin{document}
%
% \title{Contribution Title\thanks{Supported by organization x.}}
% \title{Fact-enriched fake news generation}
\title{Mining Disinformation and Fake News: Concepts, Methods, and Recent Advancements}
\titlerunning{Mining Disinformation and Fake News}
\author{Kai Shu\inst{1}, Suhang Wang\inst{2}, Dongwon Lee\inst{2}, \and Huan Liu\inst{1}}
\institute{
Computer Science and Engineering, Arizona State University, Tempe, AZ, USA
\email{\{kai.shu, huan.liu\}}@asu.edu\and
College of Information Sciences and Technology, The Penn State University, University Park, PA, USA\\
\email{\{szw494, dongwon\}}@psu.edu
}

\maketitle              % typeset the header of the contribution
\begin{abstract}

In recent years, disinformation including fake news, has became a global phenomenon due to its explosive growth, particularly on social media. The wide spread of disinformation and fake news can cause detrimental societal effects. Despite the recent progress in detecting disinformation and fake news, it is still non-trivial due to its complexity, diversity, multi-modality, and costs of fact-checking or annotation. % the diverse nature of of information domains (topics) and expensive annotation cost. 
%In addition, research on disinformation, misinformation and fake news are seemingly different defined  for the  same  problem,  and  inconsistent results in different studies. 
The goal of this chapter is to pave the way for appreciating the challenges and advancements via: (1) introducing the types of information disorder on social media and examine their differences and connections; (2) describing important and emerging tasks to combat disinformation for characterization, detection and attribution; and (3) discussing a weak supervision approach to detect disinformation with limited labeled data. We then provide an overview of the chapters in this book that represent the recent advancements in three related parts: (1) user engagements in the dissemination of information disorder; (2) techniques on detecting and mitigating disinformation; and (3) trending issues such as ethics, blockchain, clickbaits, etc.  We hope this book to be a convenient entry point for researchers, practitioners, and students to understand the problems and challenges, learn state-of-the-art solutions for their specific needs, and quickly identify new research problems in their domains. 
%We hope to enable the intensive research in this area to be conveniently reused in real-world applications and open up potential directions for future studies.

\end{abstract}

Social media has become a popular means for information seeking and news consumption. Because it has low barriers to provide and disseminate news online faster and easier through social media, large amounts of disinformation such as fake news, i.e., those news articles with intentionally false information, are produced online for a variety of purposes, ranging from financial to political gains. We take fake news as an example of disinformation. The extensive spread of fake news can have severe negative impacts on individuals and society. First, fake news can impact readers' confidence in the news ecosystem. For example, in many cases the most popular fake news has been more popular and widely spread on Facebook than mainstream news during the U.S. 2016 presidential election\footnote{https://www.buzzfeednews.com/article/craigsilverman/viral-fake-election-news-outperformed-real-news-on-facebook}. Second, fake news intentionally persuades consumers to accept biased or false beliefs for political or financial gain. For example, in 2013, \$130 billion in stock value was wiped out in a matter of minutes following an Associated Press (AP) tweet about an explosion that injured Barack Obama\footnote{https://www.telegraph.co.uk/finance/markets/10013768/Bogus-AP-tweet-about-explosion-at-the-White-House-wipes-billions-off-US-markets.html}. AP said its Twitter account was hacked. Third, fake news changes the way people interpret and respond to real news, impeding their abilities to differentiate what is true from what is not. Therefore, it is critical to understand how fake news propagate, developing data mining techniques for efficient and accurate fake news detection and intervene to mitigate the negative effects.

This book aims to bring together researchers, practitioners and social media providers for understanding propagation, improving detection  and mitigation of disinformation and fake news in social media. Next, we start with different types of information disorder.

\section{Information Disorder}\label{sec:pre}
%\subsection{Information Disorder}
Information disorder has been an important issue and attracts increasing attention in recent years.  The openness and anonymity of social media makes it convenient for users to share and exchange information, but also makes it vulnerable to nefarious activities. Though the spread of misinformation and disinformation has been studied in journalism, the openness of social networking platforms, combined with the potential for automation, facilitates the information disorder to rapidly propagate to massive numbers of people, which brings about unprecedented challenges. In general, information disorder can be categorized into three major types: disinformation, misinformation, and malinformation~\cite{wardle2017information}. \textit{Disinformation} is  fake or inaccurate information that is intentionally spread to mislead and/or deceive. \textit{Misinformation} is false content shared by a person who does not realize it is false or misleading. \textit{Malinformation} is to describe genuine information that is shared with an intent to cause harm. In addition, there are some other related types of information disorder~\cite{wu2019misinformation,zhou2018fake}: \textit{rumor} is a story circulating from person to person, of which the truth is unverified or doubtful. Rumors usually arise in the presence of ambiguous or threatening events. When its statement is proved to be false, a rumor is a type of misinformation; \textit{Urban Legend} is a fictional story that contains themes related to local popular culture. The statement and story of an urban legend are usually false. An urban legend is usually describing unusual, humorous, or horrible events; \textit{Spam} is unsolicited messages sent to a large number of recipients, containing irrelevant or inappropriate information, which is unwanted. 

The spread of false or misleading information often has a dynamic nature, causing the exchanging among different types of information disorder. On the one hand, disinformation can become misinformation.  For example, a disinformation creator can intentionally distribute the false information on social media platforms. People who see the information may be unaware that it is false and share it in their communities, using their own framing. On the other hand, misinformation can also be transformed into disinformation. For example, a piece of satire news may be intentionally distributed out of the context to mislead consumers. A typical example of disinformation is fake news. We use it as a tangible case study to illustrate the issues and challenges of mining disinformation.  
% It is worth mentioning that in real world scenarios, it is often difficult to know the intent from  

\begin{figure}[tp!]
   \centering
   \includegraphics[width=0.7\textwidth]{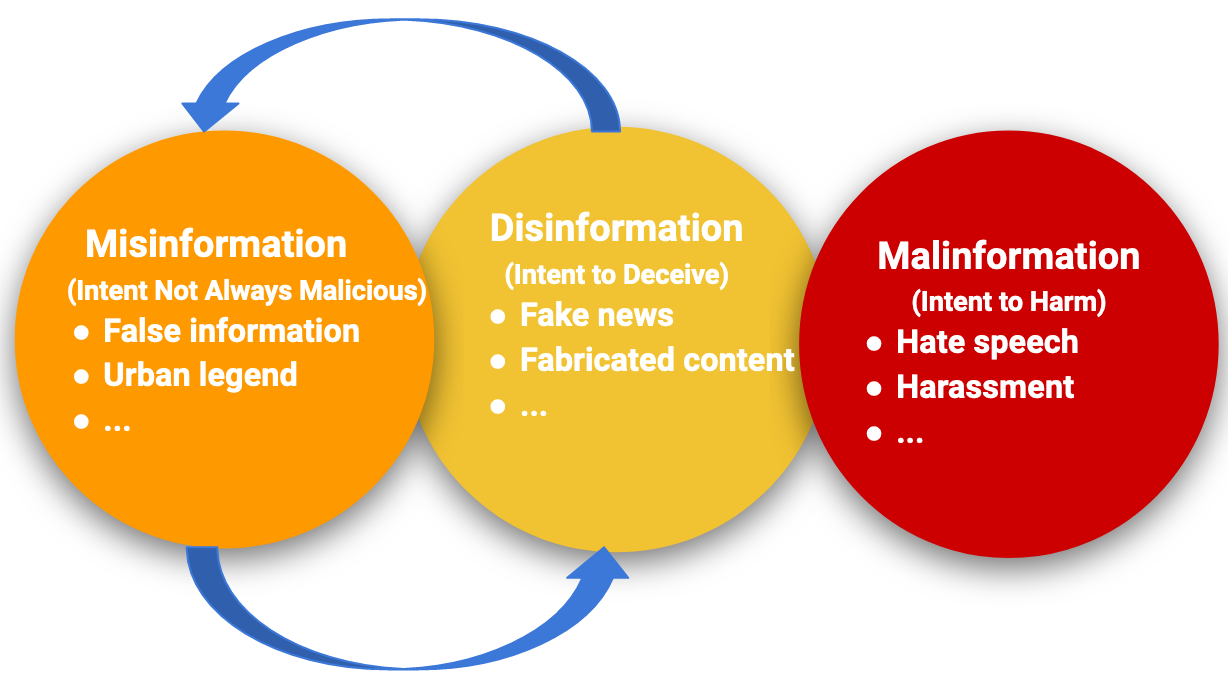}
   \caption{The illustration of the relations among disinformation, misinformation, and malinformation, with representative examples (e.g., fake news is an example of disinformation). In addition, misinformation and disinformation can be converted mutually.}
   \label{fig:concepts}
\end{figure}

% multi-modality
% language: 
% image: deep fakes

\subsection{Fake News as an Example of Disinformation} %Characterization, Detection, and Attribution}
% Social media has become a popular means to information seeking and news consuming. Because it is cheap to provide news online and much faster and easier to disseminate through social media, large volumes of disinformation such as fake news, i.e., those news articles with intentionally false information, are produced online for a variety of purposes, such as financial and political gain. 
In this subsection, we show how disinformation (fake news) can be characterized, detected, and attributed with social media data. Fake news is commonly referred as the news article that are intentionally and verifiably false and could mislead readers~\cite{tandoc2018defining,shu2019dfn}. 
% We will address other types of disinformation and misinformation based on our understanding on fake news. 

For  \textit{\textbf{characterization}}, the goal is to understand whether the information is malicious, has harmless intents, or has other insightful traits. When people create and distribute disinformation they typically have a specific purpose in mind,
or intent. For example, there can be many possible intents behind the deception including: (1) persuade people to support individuals, groups, ideas, or future actions; (2) persuade people to oppose individuals, groups, ideas or future actions; (3) produce emotional reactions (fear, anger or joy) toward some individual, group, idea or future action in the hope of promoting support or opposition; (4) educate (e. g., about vaccination threat); (5) prevent an embarrassing or criminal act from being believed; (6) exaggerate the seriousness of something said or done (e.g., use of personal email by government officials); (7) create confusion over past incidents and activities (e. g., did the U.S. really land on the moon or just in a desert on earth?); or (8) demonstrate the importance of detecting disinformation to social platforms (e. g., Elizabeth Warren and Mark Zuckerberg dispute). End to end models augmented with feature embeddings such as causal relations between claims and
evidence can be used~\cite{hidey2016identifying} to detect the intents such as Persuasive influence detection~\cite{hidey2018persuasive}. Once we have identified the intent behind a deceptive news article, we can further understand how successful this intent will be: what is the likelihood that this intent will be successful in achieving its intended purpose. We can consider measures of virality grounded in social theories to aid characterization. Social psychology points to social influence (how widely the news article has been spread) and self-influence (what preexisting knowledge a user has) as viable proxies for drivers of disinformation dissemination~\cite{zhou2018tutorial}. Greater influence from the society
and oneself skews a user’s perception and behavior to trust a news article and to unintentionally engage in its dissemination. Computational social network analysis~\cite{fakebookchapter} can be used to study how social influence affects behaviors and/or beliefs of individuals exposed to disinformation and fake news. 
% We can incorporate self and social influence to identify the intention of disinformation spreading users.

When the entire news ecosystem is considered instead of individual consumption patterns, social dynamics emerge that contribute to disinformation proliferation. According to social homophily theory, social media users tend to follow friend like-minded people and thus receive news promoting their existing narratives, resulting in an echo chamber effect. To obtain a fine-grained analysis, we can treat propagation networks in a hierarchical structure, including macro-level such as posting, reposting, and micro-level such replying~\cite{shu2020hierarchical}, which shows that structural and temporal features within information hierarchical propagation networks are statistically different between disinformation and real news. This can provide characterization complementary to a purely intent-based perspective, for instance to amplify prioritization of disinformation that may quickly have undesirable impacts after being shared with benign intent (e. g., humor) initially.

For \textit{\textbf{detection}}, the goal is to identify false information effectively, at a early stage, or with explainable factors. Since fake news attempts to spread false claims in news content, the most straightforward means of detecting it is to check the truthfulness of major claims in a news article to decide the news veracity. Fake news detection on traditional news media mainly relies on exploring news content information. News content can have multiple modalities such as text, image, video. Research has explored different approaches to learn features from single or combined modalities and build machine learning models to detect fake news. In addition to features related directly to the content of the news articles, additional social context features can be derived from the user-driven social engagements of news consumption on social media platform. Social engagements represent the news proliferation process over time, which provides useful auxiliary information to infer the veracity of news articles. Generally, there are three major aspects of the social media context that we want to represent: users, generated posts, and networks. First, fake news pieces are likely to be created and spread by non-human accounts, such as social bots or cyborgs. Thus, capturing users' profiles and behaviors by user-based features can provide useful information for fake news detection~\cite{shu2018understanding}. Second, people express their emotions or opinions toward fake news through social media posts, such as skeptical opinions and sensational reactions. Thus, it is reasonable to extract post-based features to help find potential fake news via reactions from the general public as expressed in posts. Third, users form different types of networks on social media in terms of interests, topics, and relations. Moreover, fake news dissemination processes tend to form an echo chamber cycle, highlighting the value of extracting network-based features to detect fake news. 
% multi-modality

Fake news often contains multi-modality information including text, images, videos, etc.  Thus, exploiting multi-modality information has great potentials to improve the detection performance. First, existing work focuses on extracting linguistic features such as lexical features, lexicon, sentiment and readability for binary classification, or learning neural language features with neural network structures, such as convolution neural networks (CNNs) and recurrent neural networks (RNNs)~\cite{oshikawa2018survey}. Second, visual cues are extracted mainly from visual statistical features, visual content features, and neural visual features~\cite{cao2018automatic}. Visual statistical features represent the statistics attached to fake/real news pieces. Visual content features indicate the factors describing the content of images such as clarity, coherence, diversity, etc. Neural visual features are learned through neural networks such as CNNs. In addition, recent advances aim to extract visual scene graph from images to discover common sense knowledge~\cite{bosselut2019comet}, which greatly improve structured scene graphs from visual content. 

% neural fake news

For \textit{\textbf{attribution}}, the goal is to verify the purported source or provider and the associated attribution evidence. Attribution search in social media is a new problem because social media lacks a centralized authority or mechanism that can store and certify provenance of a piece of social media data. From a network diffusion perspective, identify the provenance is to find a set of key nodes such that the information propagation is maximized~\cite{fakebookchapter}. Identifying provenance paths can indirectly find the originated provenances. The provenance paths of information are usually unknown, and for disinformation and misinformation in social media it is still an open problem. The provenance paths delineate how information propagates from the sources to other nodes along the way, including those responsible for retransmitting information through intermediaries. One can utilize the characteristics of social to trace back to the source~\cite{gundecha2013seeking}. Based on the Degree Propensity and Closeness Propensity hypotheses~\cite{barbier2013provenance}, the nodes with higher degree centralities that are closer to the nodes are more likely to be transmitters. Hence, it is estimated that top transmitters from the given set of potential provenance nodes through graph optimization. We plan to develop new algorithms which can incorporate information other than the network structure such as the node attributes and temporal information to better discover provenances.

With the success of deep learning especially deep generative models, machine-generated text can be a new type of fake news that is fluent, readable, and catchy, which brings about new attribution sources. For example, benefiting from the adversarial training, a series of language generation models are proposed such as SeqGAN~\cite{yu2017seqgan}, MaliGAN~\cite{che2017maximum}, LeakGAN~\cite{guo2018long}, MaskGAN~\cite{fedus2018maskgan}, etc. and 
unsupervised models based on Transformer~\cite{attention_is_all_you_need} using multi-task learning are proposed for language generation such as GPT-2~\cite{radford2019language} and Grover~\cite{zellers2019defending}. One important problem is to consider machine-generated synthetic text and propose solutions to differentiate which models are used to generate these text. One can perform classification on different text generation algorithms' data and explore the decision boundaries. The collections of data can be acquired from representative language generation models such as VAE, SeqGAN, TextGAN, MaliGAN, GPT-2, Grover, etc. In addition, meta-learning can be utilized to predict new text generation sources from few training examples. Moreover, some generative models such as SentiGAN~\cite{wang2018sentigan}, Ctrl~\cite{keskar2019ctrl} and PPLM~\cite{dathathri2019plug},  can generate stylized text which encodes specific styles such as emotional and catchy styles. It is important to eliminate spurious correlations in the prediction model, e. g., disentangling style factors from the synthetic text using adversarial learning, and develop prediction models with capacity to recover transferable features among different text generation models.

% \subsection{Data and Evaluation}

\section{The Power of Weak Social Supervision}\label{sec:beyond}
% some transition sentences needed

% With the rise of social media, the web has become a vibrant and lively realm where billions of individuals all around the globe interact, share, post and conduct numerous daily activities.
Social media enables users to be connected and interact with anyone, anywhere and anytime, which also allows researchers to observe human behaviors in an unprecedented  scale with new lens. User engagements over information such as news articles, including posting about, commenting on or recommending the news on social media, bear implicit judgments of the users to the news and could serve as sources of labels for disinformation and fake news detection. 

\begin{figure}[tp!]
%   \vspace{-0.3cm}
   \centering
   \includegraphics[width=0.85\textwidth]{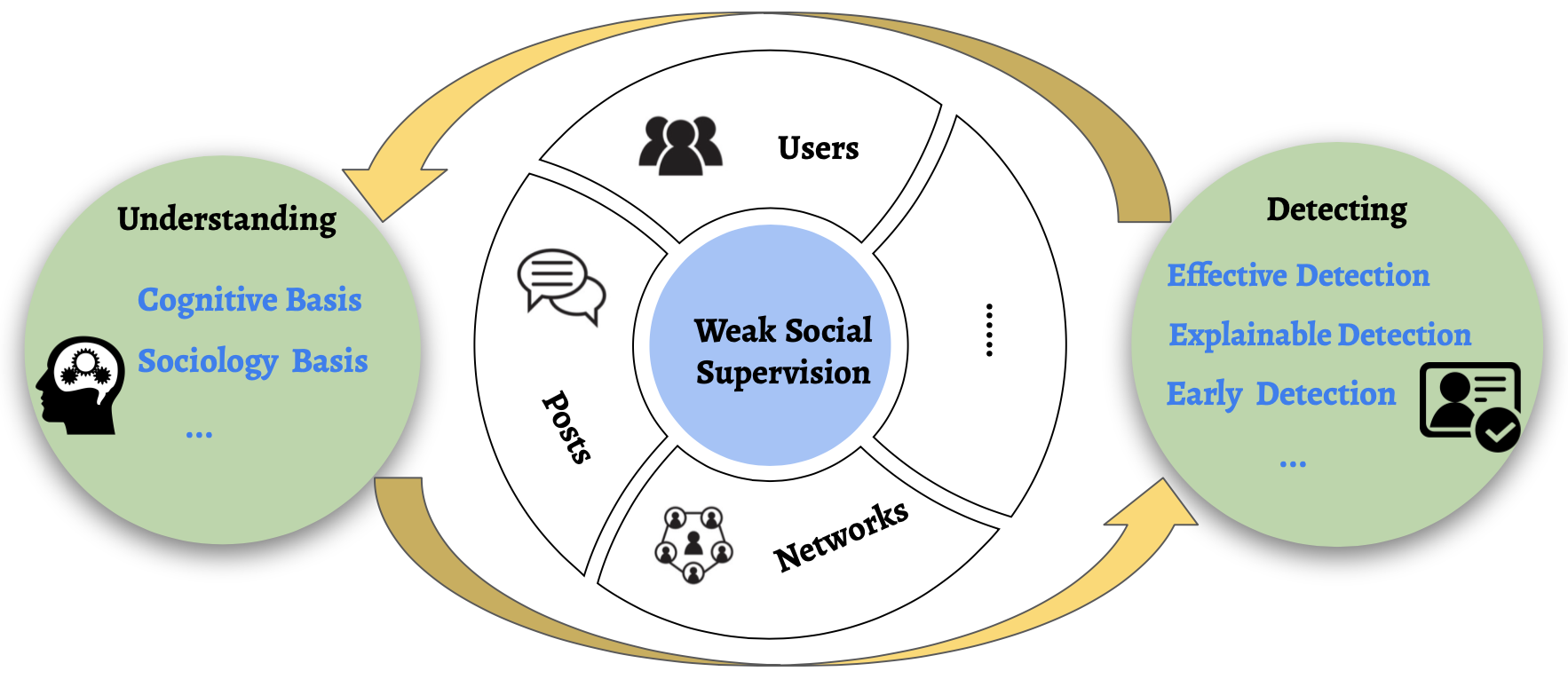}
   \caption{The illustration of learning with social supervision for understanding and detecting disinformation and fake news.}
   \label{fig:framework}
%   \vspace{-0.25cm}
\end{figure}

However, significantly different from traditional data, social media data is big, incomplete, noisy, unstructured, with abundant social relations. This new (but weak) type of data mandates new computational analysis approaches that combine social theories and statistical data mining techniques. Due to the nature of social media engagements, we term these signals as \textit{weak social supervision} (WSS).
% {\color{blue} maybe a little bit more details on what is the nature of social media engagement and why it's ``weak'' supervision}. 
We can learn with  weak social supervision to understand and detect disinformation and fake news more effectively, with explainability, at an early stage, etc. Generally, there are three major aspects of the social media engagements: users, contents, and relations (see Figure~\ref{fig:framework}). First, users exhibit different characteristics that indicate different patterns of behaviors. Second, users express their opinions and emotions through posts/comments. Third, users form different types of relations on social media through various communities. The goal of weak social supervision is to leverage signals from social media engagements to obtain weak supervision for various downstream tasks. Similar to weak supervision, we can utilize weak social supervision in the forms of weak labels and constraints. 
% {\color{blue} refer to figure 2}

% \paragraph{\textbf{Understanding disinformation with WSS}}
\subsection{Understanding disinformation with WSS}
Humans are not naturally good at differentiating misinformation and disinformation. Several cognitive theories explain this phenomenon, such as \textit{Na\"{i}ve Realism} and  \textit{Confirmation Bias}. Disinformation mainly targets consumers by exploiting the individual vulnerabilities of news consumers. With these cognitive biases, disinformation such as fake news is often perceived as real. Humans' vulnerability to fake news has been the subject of interdisciplinary research, and these results inform the creation of increasingly effective detection algorithms. To understand the influence of disinformation and fake news in social media, we can employ techniques to characterize the dissemination from various types of WSS : 1) sources (credibility/reliability, trust, stance/worldview, intentions)~\cite{abbasi2013measuring,tang2014trust}; 2) targeted social group (biases, demographic, stance/worldview)~\cite{shu2018understanding}; 3) content characteristics (linguistic, visual, contextual, emotional tone and density, length and coherence)~\cite{shu2019dfn,zhou2018tutorial}; and 4) nature of their interactions with their network (e.g., cohesive, separate)~\cite{fakebookchapter}. For example, the effects of these theories can be quantified by measuring user meta-data~\cite{shu2018understanding}, to answer the question ``why people are susceptible to fake news?'', or ``Are specific groups of people more susceptible to certain types of fake news?''.

Some social theories such as social identity theory suggests that the preference for social acceptance and affirmation is essential to a person's identity and self-esteem, making users likely to choose ``socially safe'' options when consuming and disseminating news information. 
According to social homophily theory, users on social media tend to follow and friend like-minded people and thus receive news promoting their existing narratives, resulting in an echo chamber effect. Quantitative analysis is a valuable tool for verifying whether, how, and to what magnitude these theories are predictive of user's reactions to fake news. In~\cite{shu2019hierarchical}, the authors made an attempt to demonstrate that structural and temporal perspectives within the news hierarchical propagation networks can affect fake news consumption, which indicates that additional sources of weak social supervision are valuable in the fight against fake news. To obtain a fine-grained analysis, propagation networks are treated in a hierarchical structure, including macro-level (in the form of posting, reposting) and micro-level (in the form of replying) propagation networks. It is observed that the features of hierarchical propagation networks are statistically different between fake news and real news from the structural, temporal and linguistic perspectives. 

% \paragraph{\textbf{Detecting disinformation with WSS}}
\subsection{Detecting disinformation with WSS}\label{sec:detect_wss}
Detecting disinformation and fake news poses unique challenges that makes it non-trivial. First, the \textit{data challenge} has been a major roadblock because the content of fake news and disinformation is rather diverse in terms of topics, styles and media platforms; and fake news attempts to distort truth with diverse linguistic styles while simultaneously mocking true news. Thus, obtaining annotated fake news data is non-scalable, and data-specific embedding methods are not sufficient for fake news detection with little labeled data. Second, the \textit{evolving challenge} of disinformation and fake news, meaning, fake news is usually related to newly emerging, time-critical events, which may not have been properly verified by existing knowledge bases (KB) due to the lack of corroborating evidence or claims.  To tackle these unique challenges, we can learn with \textit{weak social supervision} for detecting disinformation and fake news in different challenging scenarios such as \textit{effective}, \textit{explainable}, and \textit{early} detection strategies. The outcomes of these algorithms provide solutions to detecting fake news, also provide insights to help researchers and practitioners interpret prediction results. 

\paragraph{\textbf{Effective detection of disinformation}}
\begin{figure}[tp!]
%   \vspace{-0.3cm}
   \centering
   \includegraphics[width=0.55\textwidth]{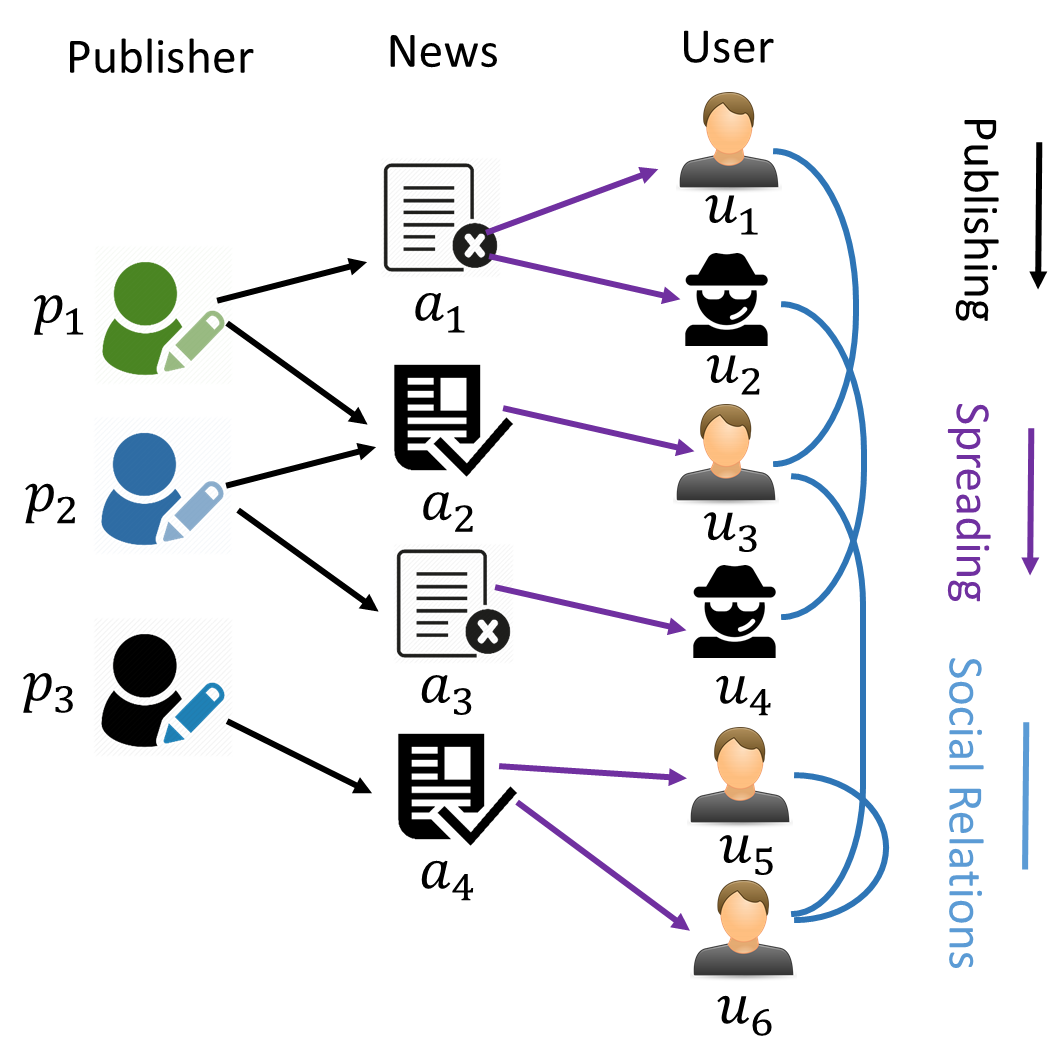}
   \caption{The TriFN model of learning with social supervision from publisher bias and user credibility for effective disinformation detection~\cite{shu2019beyond}.}
   \label{fig:interaction}
%   \vspace{-0.25cm}
\end{figure}
The goal is to leverage weak social supervision as an auxiliary information to perform disinformation detection effectively. As an example, interaction networks are used for modeling the entities and their relationships during news spreading process to detect disinformation. Interaction networks describe the relationships among different entities such as publishers, news pieces, and users (see Figure~\ref{fig:interaction}). Given the interaction networks the goal is to embed the different types of entities into the same latent space, by modeling the interactions among them. The resultant feature  representations of news can be leveraged to perform disinformation detection, with the framework \underline{Tri}-relationship for \underline{F}ake \underline{N}ews detection (TriFN)~\cite{shu2019beyond}.

Inspired from sociology and cognitive theories, the weak social supervision rules are derived. For example, social science research has demonstrated the following observations which serves our weak social supervision: \textit{people tend to form relationships with like-minded friends, rather than with users who have opposing preferences and interests.} Thus, connected users are more likely to share similar latent interests in news pieces. In addition, for publishing relationship,  the following weak social supervision can be explored: \textit{publishers with a high degree of political bias are more likely to publish disinformation.} Moreover, for the spreading relation, we have: \textit{users with low credibilities are more likely to spread disinformation, while users with high credibility scores are less likely to spread disinformation}. Techniques such as nonnegative matrix factorization (NMF) is used to learn the news representations by encoding the weak social supervision. Experiments on real world datasets demonstrate that TriFN can achieve 0.87 accuracy for detecting disinformation.

\begin{figure}[tp!]
%   \vspace{-0.3cm}
   \centering
   \includegraphics[width=0.7\textwidth]{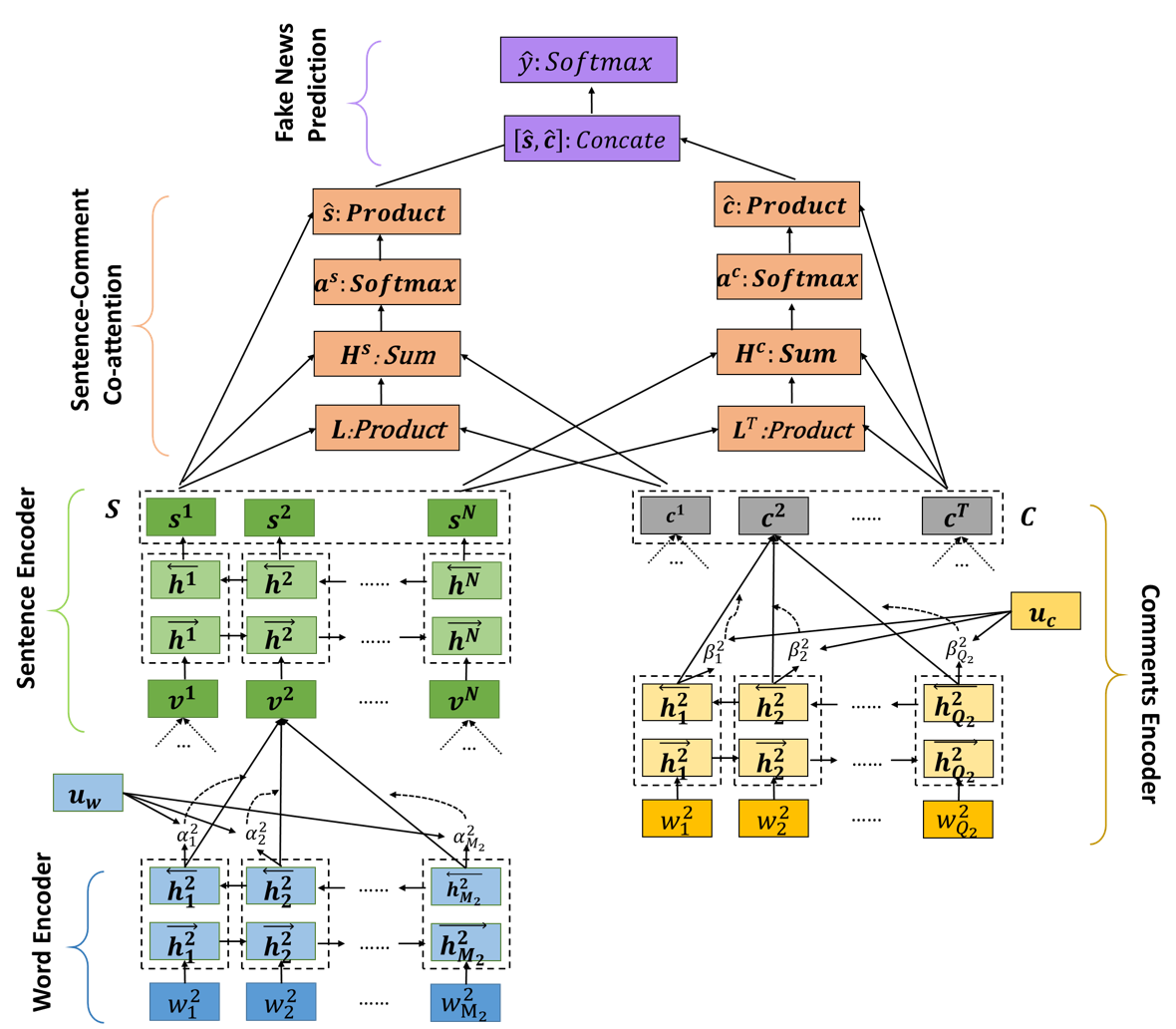}
   \caption{The dEFEND model of learning with social supervision for explainable disinformation detection~\cite{shu2019defend}.}
   \label{fig:defend}
%   \vspace{-0.25cm}
\end{figure}
\paragraph{\textbf{Confirming disinformation with explanation}}
Take fake news as an example, explainable disinformation detection aims to obtain top-$k$ explainable news sentences and user comments for disinformation detection. It has the potential to improve detection performance and the interpretability of detection results, particularly for end-users not familiar with machine learning methods. It is observed that not all sentences in news contents are fake, and in fact, many sentences are true but only for supporting wrong claim sentences. Thus, news sentences may not be equally important in determining and explaining whether a piece of news is fake or not. 
Similarly, user comments may contain relevant information about the important aspects that explain why a piece of news is fake, while they may also be less informative and noisy.  The following weak social supervision can be used: \textit{the user comments that are related to the content of original news pieces are helpful to detect fake news and explain prediction results}. In~\cite{shu2019defend}, it first uses Bidirectional LSTM with attention to learn sentence and comment representations, and then utilizes a sentence-comment co-attention neural network framework called dEFEND (see Figure~\ref{fig:defend}) to exploit both news content and user comments to jointly capture explainable factors. Experiments show that dEFEND achieves very high performances in terms of accuracy ($\sim$ 0.9) and F1 ($\sim$ 0.92). In addition, dEFEND can discover explainable comments that improve the exaplainability of the prediction results.

\paragraph{\textbf{Early warning for disinformation}}
Disinformation such as fake news is often related to newly emerging, time-critical events, which may not have been verified by existing knowledge bases or sites due to the lack of corroborating evidence. Moreover, detecting disinformation at an early stage requires the prediction models to utilize minimal information from user engagements because extensive user engagements indicate more users are already affected by disinformation. Social media data is multi-faceted, indicating multiple and heterogeneous relationships between news pieces and the spreaders on social media. First, users' posts and comments have rich crowd information including opinions, stances, and sentiment that are useful to detect fake news. Previous work has shown that conflicting sentiments among the spreaders may indicate a high probability of fake news~\cite{jin2016news,shu2017fake}. Second, different users have different credibility levels. Recent studies have shown some less-credible users  are more likely to spread fake news~\cite{shu2019beyond}. These findings  from social media have great potential to bring additional signals to early detection of fake news. Thus, we can utilize and learn with multi-source of weak social supervision simultaneously (in the form of weak labels) from social media to advance early fake news detection. 

\begin{figure}[tbp!]
	\centering
	\subfigure[MWSS Training]{
	\includegraphics[width=0.53\textwidth]{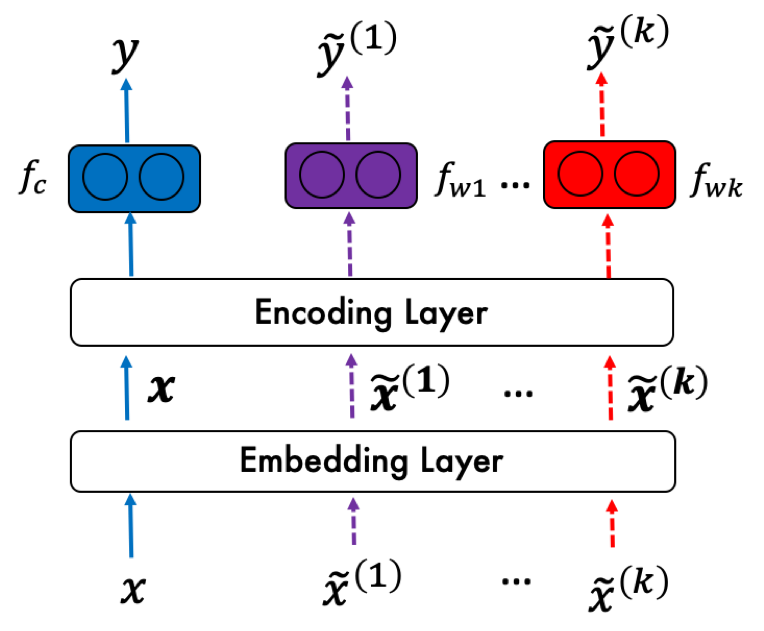}}
		\subfigure[MWSS inference]{
	\includegraphics[width=0.24\textwidth]{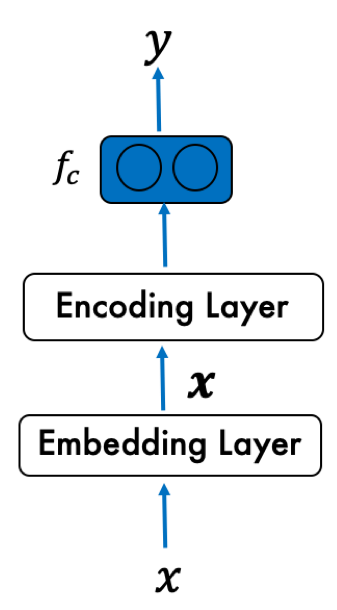}}
% 	\hspace{-0.6cm}
% 	\vspace{-0.2cm}
	\caption{The  MWSS framework for learning with multiple weak supervision from social media data for early detection of disinformation. (a) During training: it jointly learns clean labels and multiple weak sources; (b) During inference, MWSS uses the learned feature representation and function $f_c$ to predict labels for (unseen) instances in the test data. }\label{fig_wss}
% 	\vspace{-0.2cm}
\end{figure}

The key idea is that in the model training phase, social context information is used to define weak rules for obtaining weak labeled instances, in addition to the limited clean labels, to help training. In the prediction phase (as shown in Figure~\ref{fig_wss}), for any news piece in test data, only the news content is needed and no social engagements is needed at all, and thus fake news can be detected at a very early stage.
% The objective is to build a framework that leverages signals coming from multiple sources of supervision (including both clean and different weak social supervision) and learn an underlying shared representation as shown in Figure~\ref{fig_wss}. 
A deep neural network framework can be used where the lower layers of the network learn {\em shared} feature representations of the news articles, and the upper layers of the network {\em separately} model the mappings from the feature representations to each of the different sources of supervision. The framework MWSS aims to exploit jointly \underline{M}ultiple sources of \underline{W}eak \underline{S}ocial \underline{S}upervision besides the clean labels.
To extract the weal labels, the following aspects are considered including sentiment, bias, and credibility. 

First, research has shown that news with conflicting viewpoints or sentiments is more likely to be fake news~\cite{jin2016news}. Similarly, it has been shown that use opinions towards fake news have more sentiment polarity and less likely to be neutral~\cite{cui2019same}.  Thus, the sentiment scores are measured (using a widely used tool VADER~\cite{hutto2014vader}) for all the users spreading the news, and then measure the variance and desperation of the sentiment scores by computing their standard deviation. We have the following weak labeling function:
% \yl{For each rule, the threshold is set by best classifying the clean data.}
% \subsubsection{Weak Labeling Functions from User Credibility}\label{sec:weak}
\vspace{0.05in}
\begin{center}
\textit{\parbox[c]{.9\linewidth}{\textbf{Sentiment}-based:
% \yl{
% Standard deviation of users sentiments score. Sentiment score means users preference to the news. Standard deviation captures the variance and dispersion of users viewpoints towards this news. 
If a news piece has an standard deviation of user sentiment scores greater than a threshold $\tau_1$, then the news is weakly labeled as fake news. 
}}
\end{center}
\vspace{0.05in}
Second, social studies have theorized the correlation between the bias of news publisher and the veracity of news pieces~\cite{gentzkow2015media}. To some extent, the users who post the news can be a proxy of the publishers and their degree of bias toward fake news and real news is different~\cite{shu2018understanding}.  Thus, the news with users who are more biased are more likely to share fake news, and less biased users are more likely to share real news. Specifically,  method in~\cite{kulshrestha2017quantifying} is adopted to measure user bias cores by exploiting  users’ interests over her historical tweets. The bias score lies in range [-1, 1] where -1 indicates left-leaning and +1 means right-leaning. We have the following weak labeling function:
\vspace{0.05in}
\begin{center}
\textit{\parbox[c]{.9\linewidth}{\textbf{Bias}-based:
If the average value of user absolute bias scores is greater than threshold $\tau_2$, then the news pieces is weakly-labeled as fake news. }}
\end{center}
\vspace{0.05in}

Third, studies have shown that those less credible users, such as malicious accounts or normal users who are vulnerable to fake news, are more likely to spread fake news~\cite{shu2019beyond,shu2017fake}. The credibility score means the quality of being trustworthy of the user. To measure user credibility scores,we adopt the practical approach in~\cite{abbasi2013measuring}. The basic idea in~\cite{abbasi2013measuring} is that less credible users are more likely to coordinate with each other and form big clusters, while more credible users are likely to from small clusters. Note that the credibility score is inferred from users' historical contents on social media. We have the following weak labeling function:
\vspace{0.05in}
\begin{center}
\textit{\parbox[c]{.9\linewidth}{\textbf{Credibility}-based:
If a news piece has an average credibility score less than a threshold $\tau_3$, then the news is weakly-labeled as fake news.
}}
\end{center}
\vspace{0.05in}
The proper threshold is decided with a held-out validation dataset. Experimental results show that MWSS can significantly improve the fake news detection performance even with limited labeled data.

\section{Recent Advancements - An Overview of Chapter Topics}\label{sec:overview}
In this section, we demonstrate the recent advancements of mining disinformation and fake news. This book is composed of three parts, and we give an overview of the chapter topics as follows.  Part I consists of 5 chapters (2 to 6) on understanding the dissemination of information disorder. Part II contains 4 chapters (7 to 10) on techniques for detecting and mitigating disinformation, fake news, and misinformation. Part III includes 4 chapters (11 to 14) on trending issues such as ethics, block chain, clickbaits.

\vspace{0.2cm}
\noindent\textit{Part I: User Engagements in the Dissemination of Information Disorder}
% \vspace{0.2cm}
% \\
% understanding the dissemination of information disorder from user engagements
\begin{itemize}

% ``Discover Your Social Identity from What You Tweet: a Content Based Approach''
\item Understanding the characteristics of users who are likely to spread fake news is an essential step to prevent gullible users to deceived by disinformation and fake news.
In the \textbf{Chapter 2}, it presents a content-based approach to predict the social identity of users in social media. This chapter first introduces a self-attention hierarchical neural network for classifying user identities, and then demonstrates its effectiveness in a standard supervised learning setting. In addition, it shows good performance in a transfer learning setting where the framework is first trained in the coarse-grained source domain and then fine-tuned in a fine-grained target domain.

% ``User Engagement with Digital Deception''
\item User engagements related to digital threats such as misinformation, disinformation, and fake news are important dimensions for understanding and potentially defending against the wide propagation of digital threats. The \textbf{Chapter 3} performs a quantitative comparison study to showcase the user characteristics of different user groupings who engage in trustworthy information, disinformation and misinformation. It aims to answer the following questions: (1) who engage with (mis)information and (dis)information; (2) what kind of user feedback do user provide; and (3) how quickly do users engage with (mis) and (dis)information? The empirical results to these questions indicates the clear differences of  user engagement patterns, which potentially help the early handling of misinformation and disinformation.

% ``Characterization and Comparison of Russian and Chinese Disinformation Campaigns''
\item Understanding disinformation across countries is important to reveal the essential factors or players of disinformation. In the \textbf{Chapter 4}, the authors propose to identify and characterize malicious online users into multiple categories across countries of China and Russia. It first performs a comparison analysis on the differences in terms of networks, history of accounts, geography of accounts, and bot analysis. Then it explores the similarity of key actors across datasets to reveal the common characteristics of the users.

% ``Pretending Positive, Pushing False: Comparing Captain Marvel Misinformation Campaigns''
\item In the \textbf{Chapter 5}, the authors study the misinformation in the entertainment domain. This chapter compares two misinformation-fueled boycott campaigns through
examination of their origins, the actors involved, and their discussion over time.

% ``Bots, elections, and social media: a brief overview''
\item While many users on social media are legitimate, social media users may also be malicious, and in some cases are not even  real  humans.   The  low  cost  of  creating  social  media accounts  also  encourages  malicious  user  accounts,  such as social bots, cyborg users, and trolls.  The \textbf{Chapter 6} presents an overview the use of bots to manipulate the political discourse. It first illustrates the definition, creation and detection of bots. Then it uses three case studies to demonstrate the bot characteristics and engagements for information manipulation. 
\end{itemize}

\vspace{0.2cm}
\noindent\textit{Part II: Techniques on Detecting and Mitigating Disinformation}

% detecting disinformation
\begin{itemize}
% ``Tensor Embeddings for Content-based Misinformation Detection with Limited Supervision''
\item Limited labeled data is becoming the largest bottleneck for supervised learning systems. This is especially the case for many real-world tasks where large scale annotated examples can be too expensive to acquire. In the \textbf{Chapter 7}, it proposes to detect fake news and misinformation using semi-supervised learning. This chapter first presents three different tensor-based embeddings to model content-based
information of news articles which decomposition of these tensor-based models produce concise representations of spatial context.  Then it demonstrate a propagation based approach for semi-supervised classification of news articles when there is scarcity of labels.

% ``Exploring the Role of Visual Content in Fake News Detection''
\item With the development of multimedia technology, fake news attempts to utilize multimedia content with images or videos to attract and mislead consumers for rapid dissemination, which makes visual content an important part of fake news. Despite the importance of visual content, our understanding about the role of visual content in fake news detection is still limited. The \textbf{Chapter 8} presents a comprehensive review of the visual content in fake news, including the basic concepts, effective visual features, representative detection methods and challenging issues of multimedia fake news detection. It first presents an overview of different ways to extract visual features, and then discusses the models including content-based, knowledge-based approaches. This chapter can help readers to understand the role of visual content in fake news detection, and effectively utilize visual content to assist in detecting multimedia fake news.

% ``Credibility-based Fake News Detection''
\item The \textbf{Chapter 9} proposes to model credibility from various perspectives for fake news detection. First, it presents how to represent credibility from the sources of the news pieces in authors and co-authors. Then it extracts signals from news content to represent credibility including sentiments, domain expertise, argumentation, readability, characteristics, words, sentences, and typos. Finally, these credibility features are combined and utilized for fake news detection and achieve good performances in real world datasets.

% ``Standing on the Shoulders of Guardians: Novel Methodologies to Combat Fake News''
\item The intervention of misinformation and disinformation is an important task for mitigating their detrimental effects. In the \textbf{Chapter 10}, the authors propose two frameworks to intervene the spread of fake news and misinformation by increasing the guardians’ engagement in fact-checking activities. First, it demonstrates how to perform personalized recommendation of fact-checking articles to mitigate fake news. Then it tries to generate synthetic text to increase the engagement speed of fact-checking content.
\end{itemize}

% trending issues
\vspace{0.2cm}
\noindent\textit{Part III: Trending Issues}

\begin{itemize}
    \item 
% ``Developing a Model to Measure Fake News Detection Literacy of Social Media Users''
In the \textbf{Chapter 11}, it focuses on the evaluation of fake news literacy. This chapter first introduce social media information literacy (SMIL) in general. Then it applies SMIL into the context of fake news including semantic characteristics, emotional response and news sources. Finally, this chapter discusses several promising directions for both researchers and practitioners.

% ``BaitWatcher: A lightweight web interface for checking incongruent news headlines''
\item The \textbf{Chapter 12} presents a dataset, AI system, and browser extension for tackling the problem of incongruent news headlines. First, incongruent headlines are labor-intensive to annotate, the chapter proposes an automatically way to generate datasets with labels. Second, it proposes a deep neural network model that contains a hierarchical encoder to learn the representations for headlines for prediction, and demonstrate the effectiveness in real world datasets. Finally, a web interface is developed for identifying incongruent headlines in practice.

% ``An Evolving (Dis)Information Environment''
\item The \textbf{Chapter 13} presents an overview of the evolving YouTube information environment during the NATO Trident Juncture 2018 exercise, and identifies how commenters propel video’s popularity while potentially shaping human behavior through perception. This research reveals effective communication strategies that are often overlooked but highly effective to gain tempo and increase legitimacy in the overall information environment. 

% ``Blockchain Technology-based Solutions for the Problem of Misinformation: A Survey''
\item The \textbf{Chapter 14} presents a comprehensive survey on using blockchain technology to defend against misinformation. First, it gives the definition and basic concepts of blockchain. Second, it discusses how blockchain can be utilized to combat misinformation with representative approaches. Moreover, this chapter points out several promising future directions on leveraging for fighting misinformation.
\end{itemize}

\section{Looking Ahead}
Fake news and disinformation are emerging research areas and have open issues that are important but have not been addressed (or thoroughly addressed) in current studies. We briefly describe representative future directions as follows.

\paragraph{Explanatory methods} In recent years, computational detection of fake news has been producing some promising early results. However, there is a critical piece of the study, the explainability of such detection, i.e., why a particular piece of news is detected as fake. Recent approaches try to obtain explanation factors from user comments~\cite{shu2019defend} and web documents~\cite{popat2018declare}. Other types of user engagements such as user profiles can be also modeled to enhance the explainability. In addition, explaining why people are gullible to fake news and spread it is another critical task. One way to tackle this problem is from a causal discovery perspective by inferring the  directed acyclic graph (DAG) and further estimate the treatment variables of users and their spreading actions. 

\paragraph{Neural Fake News Generation and Detection} Fake news has been an important problem on social media and is amplified by the powerful deep learning models due to their power of generating neural fake news~\cite{zellers2019defending}. In terms of neural fake news generation, recent progress allows malicious users to generate fake news based on limited information.  Models like Generative Adversarial Network (GAN)~\cite{guo2018long} can generate long readable text from noise and GPT-2~\cite{radford2019language} can write news stories and fiction books with simple context. Existing fake news generation approaches may not be able to produce style-enhanced and fact-enriched text, which preserves the emotional/catchy styles and relevant topics related to news claims. Detecting these neural fake news pieces firstly requires us to understand the characteristic of these news pieces and detection difficulty. Dirk Hovy \textit{et al.} propose an adversarial setting in detecting the generated reviews~\cite{hovy2016enemy}. \cite{zellers2019defending} and~\cite{solaiman2019release} propose neural generation detectors that fine-tune classifiers on generator's previous checkpoint. It is important and interesting to explore: i) how to generate fake news with neural generative models? ii) can we differentiate human-generated and machine-generated fake/real news?

% \paragraph{\color{blue} Automatic Fake News Generation} {\color{blue} Majority of existing fake news are human written, refer to this NerurIPS paper: Defending Against Neural Fake News}

\paragraph{Early detection of disinformation} Detecting disinformation and fake news at an early stage is desired to prevent a large amount of people to be affected. Most of the previous work learns how to extract features and build machine learning models
from news content and social context to detect fake news, which generally considers the standard scenario of binary classification. More recent work consider the setting that few or even no user engagements are utilized for predicting fake news. For example, Qian \textit{et al.} propose to generate synthetic user engagements to help the detection of fake news~\cite{qian2018neural}; Wang \textit{et al.} present an event-invariant neural network model to learn transferable features to predict newly whether emerging news pieces are fake or not. We also discussed how we can utilize various types of WSS to perform early detection of fake news in Section~\ref{sec:detect_wss}. We can enhance these techniques with more sophisticated approaches that rely on less training data, for instance few-shot learning~\cite{wang2019few} for early fake news detection.

% \paragraph{Adversarial attack and defense on disinformation detection}

\paragraph{Cross topics modeling on disinformation} The content of fake news has been shown to be rather diverse in terms of topics, styles and media platforms~\cite{shu2017fake}. For a real-world fake news detection system, it is often unrealistic to obtain abundant labeled data for every domain (e.g., Entertainments and Politics are two different domains) due to the expensive labeling cost. As such, fake news detection is commonly performed in the single-domain setting, and supervised~\cite{wang2018eann} or unsupervised methods~\cite{hosseinimotlagh2018unsupervised,kai2019unsupervised} are proposed to handle limited or even unlabeled domains. However, the performance is largely limited due to overfitting on small labeled samples or without any supervision information. In addition,  models learned on one domain may be biased and might not perform well on a different target domain. One way to tackle this problem is to utilize domain adaptation techniques to explore the auxiliary information to transfer the knowledge from the source domain to the target domain. In addition, advanced machine learning strategies such as adversarial learning can be utilized to further capture the topic-invariant feature representation to better detect newly coming disinformation.

\section*{Acknowledgements}

This material is based upon work supported by, or in part by, ONR N00014-17-1-2605, N000141812108, NSF grants \#1742702, \#1820609, \#1915801. This work has been inspired by Dr. Rebecca Goolsby's vision on social bots and disinformation via interdisciplinary research.

\bibliographystyle{unsrt}
\bibliography{cite,ref_fake,acl_short}

\end{document}